\documentclass[prd,twocolumn,showpacs,floatfix,amsmath,nofootinbib,amssymb,floatfix]{revtex4}
\usepackage{graphicx,color,dcolumn,booktabs,bm}
\usepackage{longtable,lscape}
\usepackage{txfonts}
\usepackage{overpic}
\usepackage{amssymb}
\usepackage{indentfirst}
\usepackage{feynmf}   
\usepackage{slashed}  
\usepackage{cases}
\usepackage{color}
\usepackage{multirow}
\usepackage{epstopdf}
\usepackage{graphicx,color,dcolumn,booktabs,bm}
\usepackage[colorlinks,
            citecolor=blue,
            anchorcolor=red,
            menucolor=red,
            linkcolor=red,
            filecolor=red,
            runcolor=red,
            urlcolor=blue,
            frenchlinks=red]{hyperref}

\begin{document}

\title{Study on the rare decays of $Y(4630)$ induced by final state interactions }
\author{Xing-Dao Guo$^1$}
\author{Dian-Yong Chen$^2$}
\author{Hong-Wei Ke$^3$}
\author{Xiang Liu$^{4,5}$}
\author{Xue-Qian Li$^1$}
\affiliation{$^1$School of Physics, Nankai University, Tianjin
300071, P. R. China \\
$^2$Institute of Modern Physics, Chinese Academy of Sciences, Lanzhou 730000, China\\
$^3$School of Science, Tianjin University, Tianjin 300072, P.R. China\\
$^4$School of Physical Science and Technology, Lanzhou University, Lanzhou 730000, P.R. China\\
$^5$Research Center for Hadron and CSR Physics, Lanzhou University and Institute of Modern Physics of CAS, Lanzhou 730000, China     }

\vspace{2cm}
\begin{abstract}
A resonance $Y(4630)$  at the invariant mass spectrum of $\Lambda_c \bar\Lambda_c$ observed by the Belle Collaboration triggers
a hot discussion about its inner structure. Since it preferably decays into two charmed baryons $\Lambda_c\bar\Lambda_c$, it is tempted to conjecture it as
a tetraquark. Because the dominant decay portal  $Y(4630)\to \Lambda_c \bar\Lambda_c$
is close to the energy
threshold, the final state interactions may be significant and result in other baryonic and/or mesonic final states whose branching
fractions  are sizable
to be measured in the future experiments.
In this work we calculate the branching ratios of the $Y(4630)$ decays into $p\bar p$, $D^{(*)+} D^{(*)-}$,
$\pi^+\pi^-$, and $K^+K^-$ which are induced by the $\Lambda_c \bar\Lambda_c$ re-scattering.
The resultant decay patterns will be tested
by the future experiments and the consistency degree with the data composes a valuable probe for the tetraquark conjecture.
\end{abstract}

\maketitle

\section{Introduction}

The Belle Collaboration reported a charmonium-like state $Y(4630)$ in the $\Lambda_c \bar\Lambda_c$ invariant mass
spectrum from the $e^+e^-\to \Lambda_c \bar\Lambda_c$ process \cite{Pakhlova:2008vn}, where its resonance parameters
include mass $M=(4634^{+8+5}_{-7-8})$ MeV and width $\Gamma=(92^{+40+10}_{-24-21})$ MeV.  $Y(4630)$ was produced
directly by the $e^+e^-$ annihilation and its $J^{PC}$ quantum number is identified as $1^{--}$.

After the observation of $Y(4630)$, several theoretical explanations
to it were proposed (see Refs. \cite{Chen:2016qju,Liu:2013waa} for
more details). In the following, we briefly review them. In Ref.
\cite{Lee:2011rka}, authors studied the interaction of charmed
baryon and anti-charmed baryon via one boson exchange model and
explained $Y(4630)$ to be a $\Lambda_c\bar\Lambda_c$ baronium state.
Simonov proposed a model to study baryon-antibaryon production
\cite{Simonov:2011jc}, which is due to the $(q\bar{q})(q\bar{q})$
pair creation inside a hadron. By this mechanism, he further
investigated the electroproduction of $\Lambda_c \bar\Lambda_c$,
which can explain why the $Y(4630)$ enhancement structure appears in
the $\Lambda_c \bar\Lambda_c$ invariant mass spectrum
\cite{Simonov:2011jc}. There are other assumptions about the
structure that $Y(4630)$ and $Y(4660)$ may be the same state while
the later one is seen at the invariant mass spectrum of
$\psi(2S)\pi^+ \pi^-$ of the $e^+e^-$ annihilation
\cite{Wang:2007ea}, so it is naturally to assume that the state is a
molecular state of $\psi(2S)+f_0(980)$ \cite{Guo:2010tk}.
Considering $Y(4630)\to \Lambda_c \bar\Lambda_c$ and
$Y(4660)(Y(4630))\to \psi(2S)\pi^+ \pi^-$ altogether, $Y(4630)$ may
possess a small fraction of molecular component $\psi(2S)+f_0(980)$
and its decay mode can be realized via a secondary process where the
virtual $f_0(980)$ transits into a pion pair. $Y(4630)$ was
interpreted as tetraquark state in Refs.
\cite{Maiani:2014aja,Cotugno:2009ys}. Brodsky {\it et al.}
\cite{Brodsky:2014xia} proposed the color flux model to be
responsible for the diquark and anti-diquark interaction in the
tetraquark case. Maiani {\it et al.} suggested $Y(4630)$ could be a
tetraquark state with an orbital angular momentum $L=1$
\cite{Maiani:2014aja}, while Cotugno {\it et al.}
\cite{Cotugno:2009ys} indicated that $Y(4630)$ can be the first
radial excitation of another charmonium-like state $Y(4360)$ under
the tetraquark assignment.

Besides these exotic state assignments to $Y(4630)$, $Y(4630)$ were
explained as a $5^3S_1$ charmonium
\cite{Badalian:2008dv,Segovia:2008ta}. Additionally, in Ref.
\cite{vanBeveren:2008rt},  authors analyzed the experimental data of
$e^+e^-\to \Lambda_c \bar\Lambda_c$, and found that the $\Lambda_c
\bar\Lambda_c$ signal contains vector charmonia $\psi(5S) $ and
$\psi(4D)$, while the threshold behavior of $\Lambda_c\bar\Lambda_c$
cross section can be due to appearance of $\psi(3D)$
\cite{vanBeveren:2008rt}.

Although different assignments to $Y(4630)$ were given, it is
obvious that the inner structure of $Y(4630)$ is not finally
determined. Facing such research status, we still need to pay more
efforts to reveal its properties.

The key point is to explain why $Y(4630)$ was observed in the
$\Lambda_c \bar\Lambda_c$ invariant mass spectrum. Very recently,
Liu, Ke, Liu, Li \cite{Liu:2016sip} conjectured $Y(4630)$ to be a
tetraquark which is composed of a diquark and an anti-diquark, and
studied its dominate decay channel. Under this scenario, $Y(4630)$
should dominantly decay into $\Lambda_c \bar\Lambda_c$, which has
been observed in Ref. \cite{Pakhlova:2008vn}.

Along this line \cite{Liu:2016sip}, in this work we want to further
study the rare strong decay modes of $Y(4630)$. As a radially
excited state of the diquark-antidiquark bound state
\cite{Liu:2016sip}, i.e. following Cotugno {\it et al.}, Y(4630) is
supposed to be the radially-excited state of Y(4360) and $Y(4630)$
overwhelmingly decays into $\Lambda_c \bar\Lambda_c$, but since
$\Lambda_c \bar\Lambda_c$ production occurs near the threshold of
the available energy (mass of $Y(4630)$), the final state
interactions
\cite{Liu:2009iw,Liu:2006df,Liu:2006dq,Liu:2009dr,Liu:2008yy,Liu:2007ez}
at the hadron level may be significant. Such hadronic re-scattering
processes would induce a series of final products which can be
observed by future experiments even though such channels might be of
relatively small fractions. Based on the assumption that $Y(4630)$
is a pure tetraquark and the mode $Y(4630)\to \Lambda_c
\bar\Lambda_c$ dominates, we estimate the branching ratios of
several typical decay modes: $Y(4630)\to \Lambda_c \bar\Lambda_c\to$
various products, such as $p\bar p$ and $D^{(*)+} D^{(*)-}$,
$\pi^+\pi^-$ and $K^+K^-$ etc., which might be sizable for more
precise measurements.

Until now, $Y(4630)$ has only been observed in the $\Lambda_c
\bar\Lambda_c$ final state, searching for its other decay modes will
be an intriguing research topic. It is obvious that this  study will
provide a basis for further experimental exploration of $Y(4630)$.
By the measurements, the inner structure of $Y(4630)$ can be better
understood.

This paper is organized as follows. After introduction, in section
\ref{sec2}, we formulate the decay widths of $Y(4630)$ to $p\bar p$,
$D^{(*)+} D^{(*)-}$, $\pi^+ \pi^-$ and $K^+K^-$ respectively. The
numerical results are presented in the following section along with
all necessary input parameters. The last section is devoted to our
conclusion and discussions on the implications of those numerical
results.

\begin{figure*}[htbp]
\centering
\begin{tabular}{ccc}
\includegraphics[width=0.9\textwidth]{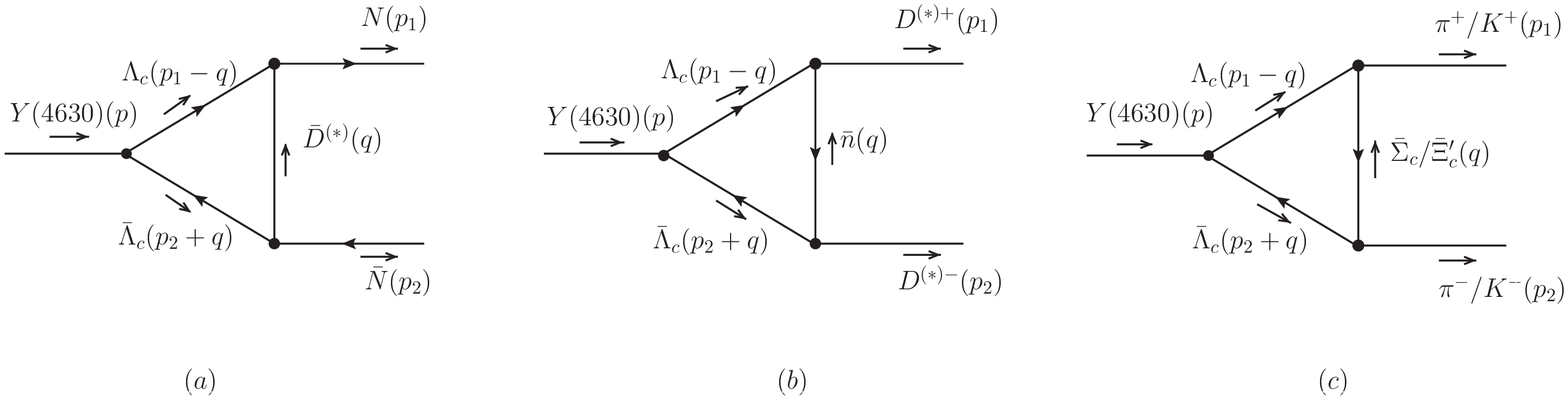}
\end{tabular}
\caption{The schematic diagrams of the decay processes for (a)
$Y(4630) \to N\bar N$, (b) $Y(4630) \to D^{(*)+} D^{(*)-}$ and (c)
$Y(4630) \to \pi^+\pi^-,\,K^+ K^-$.} \label{decay}
\end{figure*}

\section{$Y(4630)$ decays to hadron-antihadron pairs}\label{sec2}

Since only the decay mode of $Y(4630)\to\Lambda_c \bar\Lambda_c$ has
been observed so far, we assume it to be a tetraquark with hidden
charm which would preferably transit into an open-charmed
baryon-pair like $\Lambda_c \bar\Lambda_c$ \cite{Liu:2016sip}.
However, there must be other channels with smaller branching ratios
besides the dominant one, such as $e^+e^-\to Y(4630)\to N\bar N$,
$D^{(*)+} D^{(*)-}$, $\pi^+ \pi^-$ and $K^+K^-$ may occur through
re-scattering between $\Lambda_c$ and $\bar\Lambda_c$. Below, let us
focus on a few typical modes which are of sizable fractions and may
be observed in more accurate measurements. The  Feynman diagrams
related to the discussed rare strong decays are listed in Fig.
\ref{decay}: (1) The decay $Y(4630) \to N\bar N$ can occur via a
re-scattering process, where $\Lambda_c$ and $\bar\Lambda_c$
exchange a $D$ or $D^*$ meson at t-channel as shown in Fig.
\ref{decay} (a). Here, $N=p,n$ denotes nucleons. (2) The decay
$Y(4630) \to D^{(*)+} D^{(*)-}$ occurs through
$\Lambda_c-\bar\Lambda_c$ re-scattering, where a baryon is exchanged
at t-channel and the leading one should be a neutron (see Fig.
\ref{decay} (b)). (3) The decay $Y(4630) \to \pi^+ \pi^-$ $(K^+K^-)$
can occur through $\Lambda_c-\bar\Lambda_c$ re-scattering by
exchanging $\Sigma_c$  or $\Xi_c'$ baryon with spin $J=1/2$ at
t-channel. Those processes are shown in Fig. \ref{decay} (c). It
seems reasonable that our calculation should also be applicable if
the $Y(4630)$ is a molecule whose ingredients are $\Lambda_c$ and $
\bar\Lambda_c$ (so called the baryonium), since there exits strong
coupling between  $Y(4630)$ and $\Lambda_c \bar\Lambda_c$.

For quantitatively calculating these decay processes, we adopt the
effective Lagrangian approach. Here, the effective Lagrangian
depicting the interaction of $Y(4630)$ with $\Lambda_c
\bar\Lambda_c$ is \cite{Chung:1993da}
\begin{eqnarray}
\mathcal{L}_{Y(4630)\Lambda_c \bar\Lambda_c}&=&g_{Y\Lambda_c \bar\Lambda_c}Y_\mu\bar{\Lambda}_c\gamma^\mu\Lambda_c.\label{ha}
\end{eqnarray}

Following the strategy of Refs. \cite{Khodjamirian:2011jp,Khodjamirian:2011sp}, we can get the
effective interaction among $\Lambda_c$, $N$ and $D^{(*)}$ as
\begin{eqnarray}
\langle\Lambda_c(p-q)|D(-q)N(p)\rangle&=&g_{\Lambda_c N D}\bar u_{\Lambda_c}(p-q)i \gamma_5 u_N(p),\\
\langle\Lambda_c(p-q)|D^*(-q)N(p)\rangle&=&u_{\Lambda_c}(p-q)\Bigg(g_{\Lambda_c N D^*}\slashed\epsilon\nonumber\\&&+i \frac{g^T_{\Lambda_c N D^*}}{m_{\Lambda_c}+m_N}\sigma_{\mu\nu}\epsilon^\mu q^\nu\Bigg) u_N(p).
\end{eqnarray}
In the expression of $\langle\Lambda_c(p-q)|D^*(-q)N(p)\rangle$,
the second term $\frac{g^T_{\Lambda_c N D^*}}{m_{\Lambda_c}+m_N}\sigma_{\mu\nu}\epsilon^\mu q^\nu$ depends on the exchange momentum $q^\nu$
which is small, so that in practical computation it can be neglected.

In addition, $\Lambda_c$ coupling with $\Sigma_c(\Xi_c')$ and $\pi(K)$ can be expressed as
\begin{eqnarray}
\mathcal{L}_{\Lambda_c\Xi_c'K} &=& -\frac{g_2}{2f}
      \left(\bar{\Xi}_c^{'0}\gamma_{\mu}\gamma_5K^{-\mu}\Lambda_c
      -\bar{\Xi}_c^{'+}\gamma_{\mu}\gamma_5\bar{K}^{0\mu}\Lambda_c\right)+h.c.,\nonumber\\
\mathcal{L}_{\Lambda_c\Sigma_c\pi} &=& -\frac{g_2}{\sqrt{2}f}
      \Big(\bar{\Sigma}_c^+\gamma_{\mu}\gamma_5\pi^{0\mu}\Lambda_c
      -\bar{\Sigma}_c^{++}\gamma_{\mu}\gamma_5\pi^{+\mu}\Lambda_c\nonumber\\
      &&+\bar{\Sigma}_c^0\gamma_{\mu}\gamma_5\pi^{-\mu}\Lambda^+\Big)+h.c.,
\end{eqnarray}
which were constructed in Refs. \cite{Liu:2011xc,Yan:1992gz}. It
should be specially clarified that only the $SU(3)$ sextet states
$\Xi_c^\prime$ ( not the antitriplet state $\Xi_c$) appear in the
coupling, under the heavy quark limit \cite{Yan:1992gz}.

With the above preparation, we write out the decay amplitudes corresponding to Fig. \ref{decay} (a).
For the case where the exchanged meson is $D-$meson, the decay amplitude is
\begin{eqnarray}
&&\mathcal{M}_{Y(4630)\to N\bar{N}}^{D}\nonumber\\
&&=\displaystyle\int\frac{d^4q}{(2\pi)^4}\bar{u}(p_1)g_{\Lambda_c N D} i\gamma^{5}
\frac{i((\slashed{p}_1-\slashed{q})+m_{\Lambda_c})}{(p1-q)^2-m_{\Lambda_c}^2}
\nonumber\\&&\quad\times g_{Y\Lambda_c \bar\Lambda_c} \slashed{\epsilon}_Y
\frac{i(-(\slashed{p}_2+\slashed{q})+m_{\Lambda_c})}{(p_2+q)^2-m_{\Lambda_c}^2}g_{\Lambda_c N D} i\gamma^{5}{v}(p_2)\nonumber\\
&&\quad\times
\frac{1}{q^2-m_D^2}\mathcal{F}^2(m_D^2,q^2).
\end{eqnarray}
Here, $\epsilon_Y$ is the polarization vector of $Y(4630)$, $q$ denotes the exchanged momentum, $p_1(p_2)$ is the
momentum of the final state $N(\bar N$). $g_{Y\Lambda_c \bar\Lambda_c}$ and $g_{\Lambda_c ND}$ are respectively the  coupling constants
at the effective vertices $Y(4630)\Lambda_c \bar\Lambda_c$  and
$\Lambda_c ND$ whose numerical values were given in Ref. \cite{Khodjamirian:2011sp}.
If the exchanged meson is $D^*$, the amplitude is
\begin{eqnarray}
&&\mathcal{M}_{Y(4630)\to N\bar{N}}^{D^*}\nonumber\\
&&=\displaystyle\int\frac{d^4q}{(2\pi)^4}\bar{u}(p_1)g_{\Lambda_c N D^*} \gamma^{\mu}
\frac{i\left[(\slashed{p}_1-\slashed{q})+m_{\Lambda_c}\right]}{(p_1-q)^2-m_{\Lambda_c}^2}
\nonumber\\&&\quad\times g_{Y\Lambda_c \bar\Lambda_c}\slashed{\epsilon}_Y
\frac{i\left[-(\slashed{p}_2+\slashed{q})+m_{\Lambda_c}\right]}{(p_2+q)^2-m_{\Lambda_c}^2}g_{\Lambda_c N D^*} \gamma_{\mu}{v}(p_2)\nonumber\\
&&\quad\times
\frac{1}{q^2-m_{D^*}^2}\mathcal{F}^2(m_{D^*}^2,q^2).
\end{eqnarray}
In the above amplitudes, $\mathcal{F}(m_{D^{(*)}}^2,q^2)$ is the
form factor, which is introduced to compensate the off-shell effect
of exchanged $D^{(*)}$ and we will discuss it in some details in
next section.

After averaging over initial spin and summing over finial spins, the decay width $\Gamma[Y(4630) \to N\bar N]$ can be written as
\begin{eqnarray}
\Gamma\left[Y(4630) \to N\bar N\right]=\frac{|\textbf{p}|\overline{\left|\mathcal{M}(Y(4630)\to N\bar{N})\right|^2}}{24\pi m^2_{Y(4630)}}
\end{eqnarray}
with
\begin{eqnarray}
\mathcal{M}(Y(4630)\to N\bar{N})=\mathcal{M}_{Y(4630)\to N\bar{N}}^{D}+\mathcal{M}_{Y(4630)\to N\bar{N}}^{D^*},
\end{eqnarray}
where $|\textbf{p}|=\sqrt{m^2_{Y(4630)}-4m^2_N}/{2}$. In next
section we will present the numerical results of the widths which
depend on the parameter $\alpha$ which is defined below Eq.
(\ref{ff}).

In the following, we also obtain the decay amplitudes for those processes  shown in Fig. \ref{decay} (b).
For $Y(4630)\to D^+ D^-$, its amplitude is
\begin{eqnarray}
&&\mathcal{M}(Y(4630)\to D^+ D^-)\nonumber\\&&=\displaystyle\int\frac{d^4q}{(2\pi)^4}(-1){\mathrm{Tr}}\Bigg[g_{\Lambda_c N D} i\gamma^{5}
\frac{i\left[(\slashed{p}_1-\slashed{q})+m_{\Lambda_c}\right]}{(p_1-q)^2-m_{\Lambda_c}^2}\nonumber\\&& \quad\times g_{Y\Lambda_c \bar\Lambda_c}
\slashed{\epsilon}_Y
\frac{i\left[-(\slashed{p}_2+\slashed{q})+m_{\Lambda_c}\right]}{(p_2+q)^2-m_{\Lambda_c}^2}g_{\Lambda_c N D} i\gamma^{5}
\frac{i(-\slashed{q}+m_N)}{q^2-m_N^2}\Bigg]\nonumber\\&&\quad\times\mathcal{F}^2(m_N^2,q^2).
\end{eqnarray}
For the processes $Y(4630)$ decaying into $D^+ D^{*-}$ and $D^{*+} D^{*-}$, their decay amplitudes denote
\begin{eqnarray}
&&\mathcal{M}(Y(4630)\to D^+ D^{*-})\nonumber\\&&=\displaystyle\int\frac{d^4q}{(2\pi)^4}(-1)\mathrm{Tr}\Bigg[g_{\Lambda_c N D} i\gamma^{5}
\frac{i\left[(\slashed{p}_1-\slashed{q})+m_{\Lambda_c}\right]}{(p_1-q)^2-m_{\Lambda_c}^2}\nonumber\\&& \quad\times g_{Y\Lambda_c \bar\Lambda_c}
\slashed{\epsilon}_Y
\frac{i\left[-(\slashed{p}_2+\slashed{q})+m_{\Lambda_c}\right]}{(p_2+q)^2-m_{\Lambda_c}^2}g_{\Lambda_c N D^*} \slashed{\epsilon}^*_{D^*}
\frac{i(-\slashed{q}+m_N)}{q^2-m_N^2}\Bigg]\nonumber\\&&\quad\times\mathcal{F}^2(m_N^2,q^2),
\end{eqnarray}
and
\begin{eqnarray}
&&\mathcal{M}(Y(4630)\to D^{*+}D^{*-})\nonumber\\&&=\displaystyle\int\frac{d^4q}{(2\pi)^4}(-1)\mathrm{Tr}\Bigg[g_{\Lambda_c N D^\ast} \slashed{\epsilon}^*_{D^*}
\frac{i\left[(\slashed{p}_1-\slashed{q})+m_{\Lambda_c}\right]}{(p_1-q)^2-m_{\Lambda_c}^2}\nonumber\\&& \quad\times g_{Y\Lambda_c \bar\Lambda_c}
\slashed{\epsilon}_Y
\frac{i(-(\slashed{p}_2+\slashed{q})+m_{\Lambda_c})}{(p_2+q)^2-m_{\Lambda_c}^2}g_{\Lambda_c N D^*} \slashed{\epsilon}^*_{\bar D^*}
\frac{i(-\slashed{q}+m_N)}{q^2-m_N^2}\Bigg]\nonumber\\&&\quad\times\mathcal{F}^2(m_N^2,q^2),
\end{eqnarray}
respectively.
The general expression of decay width $\Gamma\left[Y(4630) \to D^{(*)+} D^{(*)-}\right]$ can be written as
\begin{eqnarray}
\Gamma\left[Y(4630) \to D^{(*)+} D^{(*)-}\right]=\frac{|\textbf{q}|\overline{\left|\mathcal{M}(Y(4630)\to D^{(*)+} D^{(*)-})\right|^2}}{24\pi m^2_{Y(4630)}},\nonumber
\end{eqnarray}
where $|\textbf{q}|=\lambda^{1/2}(m_{Y(4630)}^2,m^2_{D^{(*)}},m^2_{D^{(*)}})/(2m_{Y(4630)})$ is the three-momentum of the final states in the
center of mass frame of $Y(4630)$. $\lambda(a,b,c)=a^2+b^2+c^2-2ab-2ac-2bc$ is the K\"allen function.

Similarly, we obtain the decay amplitudes of $Y(4630)$ decaying into $\pi^+ \pi^-$ and $K^+ K^-$ as
\begin{eqnarray}
&&\mathcal{M}(Y(4630)\to \pi^+\pi^-)\nonumber\\&&=\displaystyle\int\frac{d^4q}{(2\pi)^4}(-1)\mathrm{Tr}\Bigg[\frac{g_2}{\sqrt2 f} \slashed{p}_1\gamma^{5}
\frac{i\left[(\slashed{p}_1-\slashed{q})+m_{\Lambda_c}\right]}{(p_1-q)^2-m_{\Lambda_c}^2}\nonumber\\&& \quad\times g_{Y\Lambda_c \bar\Lambda_c}
\slashed{\epsilon}_Y
\frac{i\left[-(\slashed{p}_2+\slashed{q})+m_{\Lambda_c}\right]}{(p_2+q)^2-m_{\Lambda_c}^2}\frac{g_2}{\sqrt2 f} \slashed{p}_2\gamma^{5}
\frac{i(-\slashed{q}+m_{\Sigma_c})}{q^2-m_{\Sigma_c}^2}\Bigg]\nonumber\\&&\quad\times\mathcal{F}^2(m_{\Sigma_c}^2,q^2),
\end{eqnarray}
and
\begin{eqnarray}
&&\mathcal{M}(Y(4630)\to K^+ K^-)\nonumber\\&&=\displaystyle\int\frac{d^4q}{(2\pi)^4}(-1)\mathrm{Tr}\Bigg[\frac{g_2}{2 f} \slashed{p}_1\gamma^{5}
\frac{i\left[(\slashed{p}_1-\slashed{q})+m_{\Lambda_c}\right]}{(p_1-q)^2-m_{\Lambda_c}^2}\nonumber\\&& \quad\times g_{Y\Lambda_c \bar\Lambda_c}
\slashed{\epsilon}_Y
\frac{i\left[-(\slashed{p}_2+\slashed{q})+m_{\Lambda_c}\right]}{(p_2+q)^2-m_{\Lambda_c}^2}\frac{g_2}{2 f} \slashed{p}_2\gamma^{5}
\frac{i(-\slashed{q}+m_{\Xi_c'})}{q^2-m_{\Xi_c'}^2}\Bigg]\nonumber\\&&\quad\times\mathcal{F}^2(m_{\Xi_c'}^2,q^2).
\end{eqnarray}

%

\section{The numerical results\label{sec3}}

Now in terms of the formulation derived in the past section, we numerically compute the decay widths of the aforementioned processes. The
input parameters are taken from Refs. \cite{Pakhlova:2008vn,Liu:2011xc,Agashe:2014kda}, totally, we have $g_2=0.598$, $f=92.3$ MeV,
$m_{Y(4630)}=4.630$ GeV, $m_{D}=1.865$ GeV, $m_{D^*}=2.007$ GeV, $m_{\Lambda_c}=2.286$ GeV, $m_{\Sigma_c}=2.455$ GeV, $m_{\Xi_c'}=2.578$ GeV,
$m_{K}=0.497$ GeV, $m_{\pi}=0.135$ GeV, $m_{n}=0.940$ GeV and $m_p=0.938$ GeV. The coupling constants
$g_{\Lambda_c N D}=10.7^{+5.3}_{-4.3}$ and $g_{\Lambda_c N D^*}=-5.8^{+2.1}_{-2.5}$ are borrowed from Refs. \cite{Khodjamirian:2011jp,Khodjamirian:2011sp},
where we take those central values in our calculation.
The coupling constant $g_{Y\Lambda_c \bar\Lambda_c}$ is obtained by fitting the available experimental data.
Since the branching ratio of $\mathcal{B}(Y(4630)\to \Lambda_c \bar\Lambda_c)$ is not very accurately determined yet and the resonance peak is only observed at this channel,
we have a reason to assume that $Y(4630)$
predominantly decays into $\Lambda_c \bar\Lambda_c$, so its partial width is approximately equal to the total width of $Y(4630)$. With the effective Lagrangian in Eq. (\ref{ha}),
the decay width is written as
\begin{eqnarray}
\Gamma(Y(4630) \to \Lambda_c \bar\Lambda_c)=\frac{|\textbf{k}|(m^2_{Y(4630)}+2m^2_{\Lambda_c})}{6\pi m^2_{Y(4630)}}
g_{Y\Lambda_c \bar\Lambda_c}^2,
\end{eqnarray}
where {$|\textbf{k}|=\sqrt{{m^2_{Y(4630)}-4m^2_{\Lambda_c}}}/2$}. By fitting the experimental width of $Y(4630)$ ($\Gamma_{Y(4630)}=92$ MeV
\cite{Pakhlova:2008vn}), where we assume $Y(4630) \to \Lambda_c \bar\Lambda_c$ to be dominate decay of $Y(4630)$, we obtain $g_{Y\Lambda_c \bar\Lambda_c}=1.78$.


The form factor at the effective hadronic vertices is introduced to compensate the off-shell
effects of the intermediate agents (baryon or meson), and a reasonable choice for it is suggested by
Cheng {\it et al.} \cite{Cheng:2004ru} as
\begin{equation}
\mathcal{F}(m_E^2,q^2)=\frac{\Lambda^2-m_E^2}{\Lambda^2-q^2},
\label{ff}
\end{equation}
where  the cut-off parameter $\Lambda$ can be
parametrized as  $\Lambda=\alpha \Lambda_{QCD}+m_E$ with $\Lambda_{QCD}=220$ MeV and $\alpha$ 
is a phenomenological parameter of order of unity \cite{Cheng:2004ru}. In the above expression, $m_E$ denotes the mass of the exchanged particle.

\begin{center}
\renewcommand\arraystretch{1.4}
\begin{table}[htbp]
\begin{tabular}{cccc}
\toprule[1pt]\toprule[1pt]
Channel&\multicolumn{2}{c}{Branching ratio} \\
\midrule[1pt]
                  &$\alpha=1.5$                 &$\alpha=1.7$&\\
$D^+ D^-$   &  0.085  (fixed) \cite{Pakhlova:2008zza} &0.14 (fixed) \cite{Aubert:2006mi}\\
$D^+ D^{*-}+h.c.$& 0.122                    & 0.193                     &\\
$D^{*+} D^{*-}$  &  0.094                    &  0.146                  &\\
$p\bar{p}$      &  0.037                    & 0.062                    &\\
$\pi^+\pi^-$   &  $1.65\times 10^{-6}$      & $2.62\times 10^{-6}$         &\\
$K^+K^-$       &  $3.63\times 10^{-6}$     & $5.73\times 10^{-6}$        &\\
\bottomrule[1pt]\bottomrule[1pt]
\end{tabular}
\caption{The calculated upper limit for the branching ratios of $Y(4630) \to D^{(*)+} D^{(*)-}$, $p\bar p$, $\pi^+\pi^-$ and $K^+ K^-$ with typical
values $\alpha=1.5$ and $1.7$. 
extracted by the similar way to that of $Y(4630)\to D^+ D^-$. 
\cite{Abe:2006fj} and $\sigma(e^+e^-\to D^{*+} D^{*-})=0.44\pm0.12$ nb at $\sqrt{s}=4.63$ GeV \cite{Abe:2006fj} were applied to this estimate.
\label{ratio1}}
\end{table}
\end{center}

\begin{figure}[htp]
\centering
\includegraphics[width=0.50\textwidth]{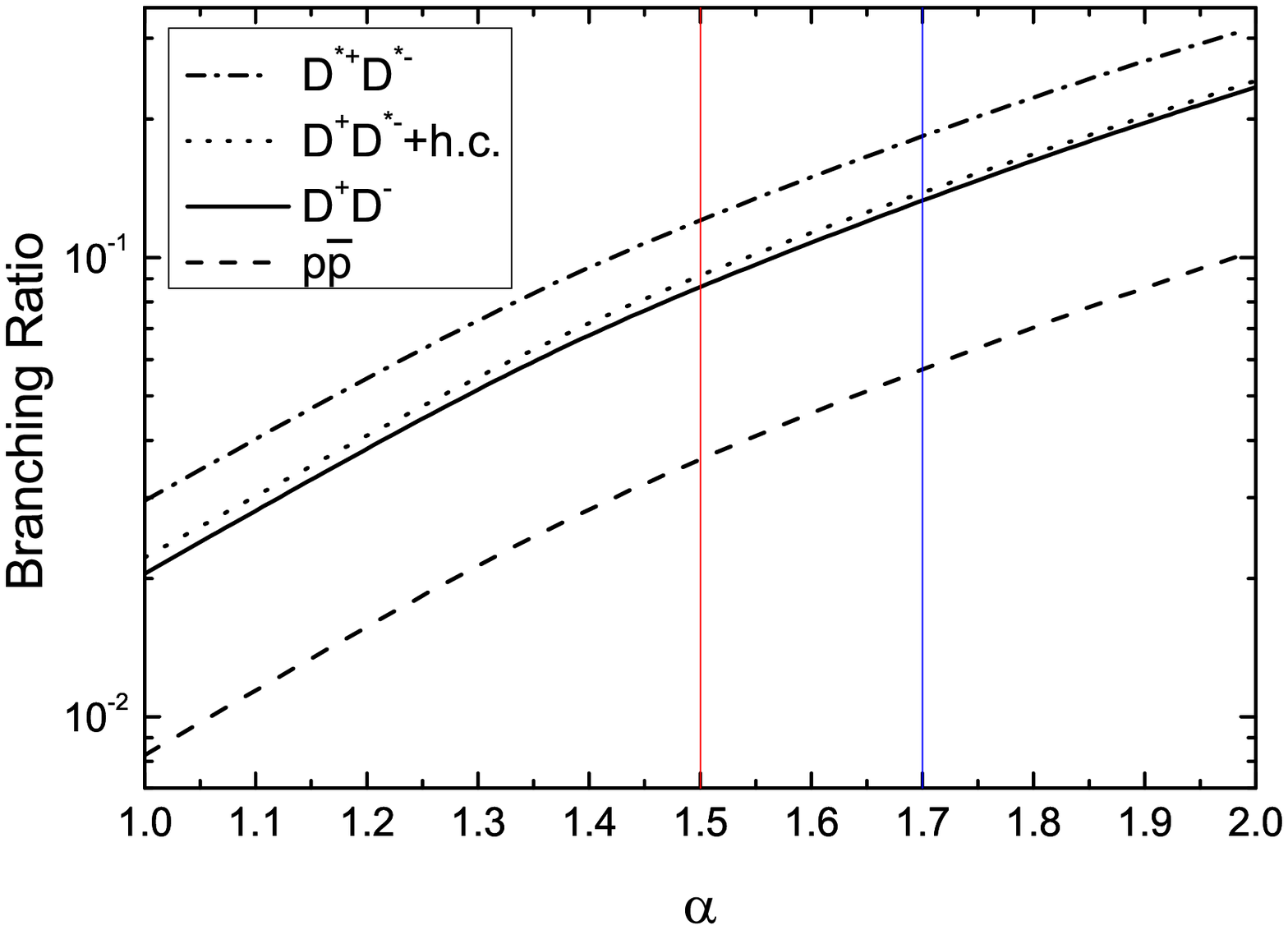}
\includegraphics[width=0.50\textwidth]{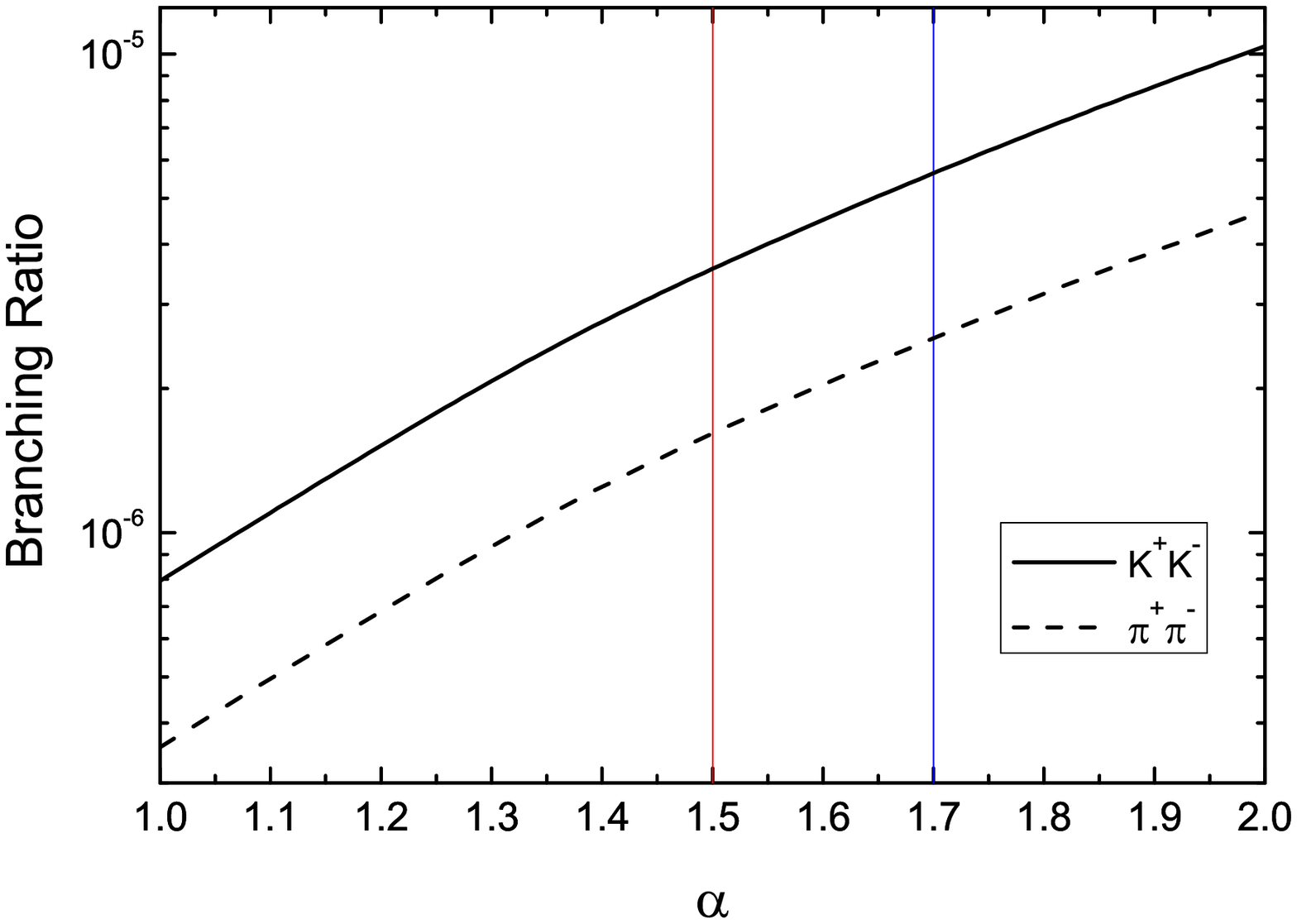}
\caption{The upper limit for the branching ratios of the rare strong decays of $Y(4630)$ dependent on $\alpha$. Here, the red and blue vertical lines correspond to $\alpha=1.5$ and 1.7, respectively.}
\label{alpha}
\end{figure}

In the following, we discuss how to constrain the $\alpha$ value by the experimental data.
We note that the cross section of $e^+e^-$ annihilation into $D^+ D^-$ has been measured by Belle \cite{Pakhlova:2008zza} and BaBar \cite{Aubert:2006mi},
while the cross section for $e^+e^-\to \Lambda_c \bar\Lambda_c$ has been given by the Belle collaboration \cite{Pakhlova:2008vn}. {The general expression of cross sections for the $e^+ e^-\to Y(4630) \to f$ is
\begin{eqnarray}
\sigma(e^+ e^- \to Y(4630) \to f)= \frac{12 \pi \Gamma_{Y}^{e^+ e^-} \mathcal{B}(Y \to f) \Gamma_Y}{(s-m_Y^2)^2+m_Y^2 \Gamma_Y^2},
\end{eqnarray} 
where $f$ denotes the final states, $\Gamma_{Y}^{e^+ e^-}$ and $\mathcal{B}(Y\to f)$ are the dileption partial width of the $Y(4630)$ and the branching ratio for the $Y(4630) \to f$ decay, respectively. Thus, the ratio of the partial widths for $D^+ D^-$ and $\Lambda_c \bar{\Lambda}_c$ can be defined as
\begin{eqnarray}
R_{D^+ D^-} &\equiv& \frac{\Gamma\left[Y(4630) \to D^+ D^-\right]}{\Gamma\left[Y(4630) \to \Lambda_c \bar\Lambda_c\right]}\nonumber\\
&=&\frac{\sigma(e^+e^- \to Y(4630) \to D^+ D^-)}{\sigma(e^+e^- \to Y(4630) \to \Lambda_c \bar\Lambda_c)}.
\end{eqnarray}
The signal of the $Y(4630)$ in the $e^+ e^- \to \Lambda_c \bar{\Lambda}_c$ had been clearly observed and the cross section for $e^+ e^-\to Y(4630) \to \Lambda_c \bar{\Lambda}_c$ was reported to be $0.47^{+0.22}_{-0.23}$ nb at $\sqrt{s}=4.630$ GeV \cite{Pakhlova:2008vn}. However, in the cross sections for $e^+ e^- \to D^+ D^-$, the signal of the $Y(4630)$ has not been observed \cite{Aubert:2006mi,Pakhlova:2008zza}. Here, we suppose that only some fraction of the $e^+e^-\to D^+ D^-$ cross section at 4.63
GeV turns out to be due to the Y(4630). Then, we take $\sigma(e^+ e^- \to Y(4630) \to D^+ D^-)= g \sigma(e^+ e^- \to D^+ D^-)$ with $g \leq 1$. According to $\sigma(e^+e^- \to Y(4630) \to D^+ D^-)=0.04\pm0.035$ nb and $0.065\pm0.055$ nb from the Belle \cite{Pakhlova:2008zza} and BaBar collaborations
\cite{Aubert:2006mi}, respectively, we get $R_{D^+ D^-}=g(0.085^{+0.024}_{-0.064})$ and $g(0.14^{+0.034}_{-0.098})$, which indicates that the upper limit of the $R_{D^+D^-}$ is determined to be $0.085^{+0.024}_{-0.064}$ or $0.14^{+0.034}_{-0.098}$ depending on the data given by different collaborations. With the upper limit of the $R_{D^+ D^-}$, we can fix the parameter $\alpha$ introduced for our model dependent calculations, then with the determined $\alpha$ we can roughly estimate the upper limits of the branching ratios of other rare decays. In addition, one should notice that the experimental
errors of the cross sections of  $e^+e^- \to D^+ D^-$ are relatively
large at the central values of the cross sections around 4.6 GeV,
thus further precise measurements in this region should provide us
more information of the $Y(4630)$ resonance. 

Under the diaquark-antidiquark
assignment to $Y(4630)$ suggested in Ref. \cite{Liu:2016sip}, we
take the width of $Y(4630)$ as the input of $\Gamma\left[Y(4630) \to
\Lambda_c \bar\Lambda_c\right]$. Then, we estimate
$\Gamma\left[Y(4630) \to D^+ D^-\right]=7.8^{+6.7}_{-6.5}$ MeV and
$12.8\pm10.3$ MeV, by which we fix $\alpha=1.5$ and $1.7$,
respectively, by the formula of $Y(4630)\to D^+ D^-$ presented in
Sec. \ref{sec2}. In the following, we adopt the obtained typical
$\alpha=1.5$ and $1.7$ to further estimate other rare strong decays
of $Y(4630)$ discussed in this work, which are shown in Table
\ref{ratio1}.

}

We list the branching ratios of the rare strong decays of $Y(4630)$
with typical $\alpha$ values in Table \ref{ratio1}, and will discuss
dependence of these branching ratios of the rare strong decays of
$Y(4630)$ on $\alpha$ (see Fig. \ref{alpha}).

Fig. \ref{decay} (a) and (c) demonstrate that there exists an OZI
suppression for $c \bar c$ annihilation while for the K meson
production (Fig. \ref{decay} (c))  $s \bar s$ annihilation is also
OZI suppressed. Thus, one expects that the branching ratios
determined by Fig. \ref{decay} (a) and (c) are somewhat smaller than
that of Fig. \ref{decay} (b). As shown in Table \ref{ratio1}, the
obtained results reflect this fact.

\section{discussion and conclusion}

In this work, based on the postulation that $Y(4630)$ observed by
the Belle collaboration at the invariant mass spectrum of $\Lambda_c
\bar\Lambda_c$ and not at other channels, is a tetraquark composed
of a diquark and an anti-diquark \cite{Liu:2016sip}, we suggest that
it overwhelmingly decays into $\Lambda_c \bar\Lambda_c$. Since the
production of $\Lambda_c \bar\Lambda_c$ is close to the available
energy threshold, the hadronic final state interactions might be
significant. The inelastic re-scattering processes may produce other
final state particles which can be either mesons or baryons.

By the standard strategy for dealing with inelastic re-scattering
processes, we provide several predictions on the branching ratios of
$Y(4630)$ decays into $p\bar{p}$, $D^{(*)+} D^{(*)-}$, $\pi^+\pi^-$,
and $K^+K^-$. As the free parameter in our calculation, $\alpha$ is
fixed by fitting the experimental data of
\cite{Aubert:2006mi,Pakhlova:2008zza} whose procedures is just
illustrated in Sec. \ref{sec3}. As a matter of fact, some other
channels, such as vector meson pairs $\rho\rho$ etc. may also exist
in the final states with similar orders of magnitude. Definitely, we
do not cover all of them, but select several typical processes to
show the scenario.

Since we take the total width of $Y(4630)$ as the partial width of
$Y(4630)\to\Lambda_c \bar\Lambda_c$ for our numerical computations,
certain errors might be caused. However, as we argued above, the
$Y(4630)\to\Lambda_c \bar\Lambda_c$ is the overwhelming mode, the
errors brought up by the approximation are not serious and the
subsequent results are trustworthy, in particular the quantitative
conclusion should not be changed.

The suggested final states are of smaller branching ratios as
expected,  our numerical results show that they are at order
$\mathcal O(10^{-3})$ to $\mathcal O(10^{-2})$, which are too small
to be detected by the present experiments, but will definitely be
"seen" by the future much more precise measurements.

Obviously, if $Y(4630)$ is not a tetraquark, but a molecular state
as suggested by some authors, or their mixture, its decay pattern
would be different from our prediction based on the tetraquark
assumption. Namely, if $Y(4630)$ is a molecular state with more
components beside $\Lambda_c \bar\Lambda_c$, one would expect other
final states to have larger branching fractions than we estimate in
this work; while for the ¡°dynamical¡± diquark model of Brodsky {\it
et al.} \cite{Brodsky:2014xia} in which the diquarks are far
separated, since the two components are far apart, except $\Lambda_c
\bar\Lambda_c$ no other direct final states could be produced.
Therefore  future measurements on various decay channels of
$Y(4630)$ will help to pin down the identity of this resonance.


\section*{Acknowledgments}
We would like to thank the anonymous referee for his/her suggestive comments.
This  work is supported by National Natural Science Foundation of China under the
Grant No. 11375128, No. 1135009, No. 11222547, No. 11175073 and No. 11375240. Xiang Liu is also supported by the National Youth
Top-notch Talent Support Program ("Thousands-of-Talents
Scheme").


\end{document}